\documentclass[]{pasj02} 
\usepackage[switch,mathlines]{lineno} 
\usepackage{url}
\usepackage{lscape}

\jyear{2025}
\Received{}
\Accepted{2025/09/18}

\begin{document} 

\title{XRISM High-resolution Spectroscopy of SS 433: Evidence of Decreasing Line-of-Sight Velocity Dispersion along the Jet}

\author{
Megumi \textsc{Shidatsu}\altaffilmark{1,2}\orcid{0000-0001-8195-6546}
\thanks{Corresponding Authors: Megumi Shidatsu, Shogo Kobayashi}}
\email{shidatsu.megumi.wr@ehime-u.ac.jp,shogo.kobayashi@rs.tus.ac.jp} 
\author{Shogo \textsc{Kobayashi}\altaffilmark{3}\orcid{0000-0001-7773-9266}}
\author{Yusuke \textsc{Sakai}\altaffilmark{4}\orcid{0000-0002-5809-3516}}
\author{Toshihiro \textsc{Takagi}\altaffilmark{1}}
\author{Yuta \textsc{Okada}\altaffilmark{5}}
\author{Shinya \textsc{Yamada}\altaffilmark{4}\orcid{0000-0003-4808-893X}}
\author{Yoshihiro \textsc{Ueda}\altaffilmark{5}\orcid{0000-0001-7821-6715}}
\author{Hideki \textsc{Uchiyama}\altaffilmark{6}\orcid{0000-0003-4580-4021}}
\author{Robert \textsc{Petre}\altaffilmark{7}\orcid{0000-0003-3850-2041}
}


\altaffiltext{1}{Graduate School of Science and Engineering, Ehime University, 2-5 Bunkyo-cho, Matsuyama, Ehime 790-8577, Japan}
\altaffiltext{2}{Research Center for Space and Cosmic Evolution, Premier Institute for Advanced Studies, Ehime University, 2-5 Bunkyo-cho, Matsuyama, Ehime 790-8577, Japan}
\altaffiltext{3}{Faculty of Physics, Tokyo University of Science, 1-3 Kagurazaka, Shinjuku-ku, Tokyo 162-8601, Japan}
\altaffiltext{4}{Department of Physics, Rikkyo University, 3-34-1 Nishi Ikebukuro, Toshima-ku, Tokyo 171-8501, Japan}
\altaffiltext{5}{Department of Astronomy, Kyoto University, Kitashirakawa-Oiwake-cho, Sakyo-ku, Kyoto 606-8502, Japan}
\altaffiltext{6}{Faculty of Education, Shizuoka University, 836 Ohya, Suruga-ku, Shizuoka, Shizuoka 422-8529, Japan}
\altaffiltext{7}{NASA/Goddard Space Flight Center, Greenbelt, MD 20771, USA}



\maketitle

\begin{abstract} 
We report on the jet structure in SS 433 based on X-ray high resolution spectroscopy with the XRISM/Resolve. The source was observed over 5 days covering both inside and outside an eclipse of the compact object by the companion star. Doppler-shifted, ionized Fe and Ni K emission lines were resolved, as well as lower-energy lines including Si and S K lines.
Time-resolved spectral analysis showed that Fe and Ni K line widths were $1020 \pm 40$ km s$^{-1}$ (corresponding the 1$\sigma$ width) in the eclipse phase, gradually increased during the egress, and reached $1740 \pm 30$ km s$^{-1}$ outside the eclipse. 
A time-averaged spectrum outside the eclipse confirmed that the Fe and Ni K lines in 5.5--9 keV are significantly broader than the Si and S K$\alpha$ emission lines in 2--4 keV. Specifically, the width in 5.5--9 keV was measured to be $1900 \pm 80$ km s$^{-1}$, whereas the width in 2--4 keV is $1300^{+300}_{-400}$ km s$^{-1}$ for the approaching (blueshifted) jet component. These results indicate that radial velocity dispersion of the jet plasma in SS 433 decreases as it moves outward. We interpret this variation as progressive jet collimation along its axis, as suggested by \citet{namiki2003}, or a decrease in turbulence in the jet plasma flow within the X-ray emitting region. We also detected a clear difference in velocity dispersion between the approaching and receding (redshifted) jet components in the 5.5--9 keV band outside eclipse. The receding jet exhibited a smaller velocity dispersion ($1400 \pm 200$ km s$^{-1}$) than the approaching jet. Since the observation was conducted when the approaching jet was tilted toward the observer, this may suggest that the receding jet was more extensively occulted by the accretion disk.

\end{abstract}


\section{Introduction}

SS 433 has been known as an unusual Galactic X-ray binary that persistently shows bipolar extended baryonic jets (see e.g., \cite{fabrika2004} for reviews). Its spectrum exhibits highly ionized blueshifted and redshifted emission lines from approaching and receding jets, respectively, viewed at high inclination with respect to the accretion disk plane. 
Unlike other typical X-ray binaries, the observed X-ray spectrum of SS 433 is dominated by thin thermal plasma emission from the jets, instead of the direct disk emission and its Comptonisation. 
Many observations have been conducted so far in various wavelengths to determine its binary parameters. 
It consists of a compact object and an A-type giant companion star, orbiting with a period of 13.082 days \citep{goranskij2011}. The nature of the compact object is one of the most intensively studied issues in this system. Various mass estimates have been reported, with some supporting a neutron star and others favoring a black hole (e.g., \cite{gies02, hillwig04, kubota10, robinson17}), although recent studies support a black hole interpretation from the long-term variation of the orbital period (\cite{cherepashchuk19, cherepashchuk21}). The jets are known to be ejected at a velocity of $0.26 c$ and show a precession with a period of $\sim$ 162 days (e.g., \cite{abell79,margon89}). Due to the high inclination angle ($78.8^\circ$; \cite{margon89}), SS 433 undergoes an eclipse of the compact object and surrounding structures by the companion star in each orbit. During the eclipse periods, the companion star obscures the innermost X-ray emitting region of the jets, resulting in a softer X-ray spectrum (e.g., \cite{kawai89,marshall2013}). 

It is also the only Galactic source with steady super-Eddington accretion, 
shining at $\gtrsim 10^{40}$ erg/s in the UV band (\cite{cherepashchuk82,dolan97}), whereas its X-ray luminosity is only $10^{35}$--$10^{36}$ erg/s (e.g., \cite{brinkmann91}). This is reminiscent of the ultraluminous supersoft sources (ULSs) in nearby galaxies, which are as luminous as ultraluminous X-ray sources (ULXs) but have very soft thermal spectra with a temperature of $\lesssim 0.1$ keV. Indeed, one of the ULSs was found to have a relativistic baryonic jet \citep{liu2015} like SS 433. In X-rays, \citet{middleton2021} suggested using NuSTAR data that the intrinsic X-ray luminosity of SS 433 exceeds $> 10^{39}$ erg/s, which is comparable to ULXs, when the system is viewed in the face-on direction. The jets were also estimated to carry kinetic power comparable to, or maybe exceeding, the Eddington luminosity, on the order of $ 10^{38}$--$10^{39}$ erg s$^{-1}$ \citep{panferov1997, kotani1998, marshall2002}.
Because of its proximity, SS 433 is a unique and ideal target 
to study the structure of jets launched at such extreme mass accretion rates.

The jet plasma is expected to cool as it propagates outward, resulting in emission at progressively lower energies. The X-ray emission from the jets is therefore considered a valuable probe of the regions closest to the jet base. 
Using a Chandra HETG spectrum of SS 433, \citet{namiki2003} found that emission lines at higher energies (Fe and Ni) are broader than lower energy lines (Si and S), indicating the decrease of the radial velocity dispersion along the jet axis. \citet{namiki2003} interpreted it as progressive jet collimation (a decrease of the jet opening angle) at a distance of $\sim 10^{12}$ cm from the compact object. 
However, the significance of their result has been controversial. Due to the limited energy resolution and statistics, the above conclusion was only derived by fitting several different lines with different ionization stages with empirical Gaussian models assuming the same widths. 
As \citet{marshall2013} pointed out, the Fe XXV He$\alpha$ line at 6.7 keV, which is the strongest line above 5 keV and mainly determines the velocity dispersion of the inner X-ray jets in \citet{namiki2003},
could be contaminated by several unresolved lines and thus cause overestimation of the line width. It is therefore essential to resolve individual lines and measure the line width 
accurately. Achieving this and establishing the true structure of the jet require spectra with higher energy resolution and better statistics. 

In this article, we tackle this issue using X-Ray Imaging and Spectroscopy Mission (XRISM; \cite{tashiro20}), which carries the X-ray microcalorimeter named Resolve \citep{ishisaki22}, with unprecedented energy resolution and with a good photon-collecting power, especially around the Fe K band. It enables us to resolve the Fe K lines from the jet in SS 433 and to measure the line widths accurately. In addition to comparing different lines in different energy bands in the same manner as \citet{namiki2003}, we study the time variation of the line widths in the Fe K band during and outside an eclipse. If the jets are progressively collimated in the line emitting region, the line widths should be larger during a non-eclipse period than an eclipse period, in which the inner part of the jets are hidden. 

This article is structured as follows. Section~\ref{sec:obs} gives the details of the XRISM observation and reduction of the Resolve data. Then, we describe our analysis and results in Section~\ref{sec:ana}. 
Finally we discuss the results and draw conclusions of this study in Section~\ref{sec:discussion}. Throughout the article, the errors represent 90\% confidence ranges for one parameter, unless otherwise specified.

\section{Observation and Data Reduction} \label{sec:obs}
The XRISM observation was performed from 2024 April 10 UT 13:41 to 15 UT 10:32 (OBSID$=$300041010), with a total exposure of $\sim$ 200 ks. In this observation, the X-ray microcalorimeter Resolve was operated under the closed gate valve condition, without any additional filters, 
and the X-ray CCD Xtend \citep{mori22,noda25, uchida25} was in the full-window mode. 
According to \citet{goranskij2011}, the period of our observation corresponds to the orbital phase of $\phi_{\rm orb}=$ 0.003--0.375 (where phase 0 corresponds to the inferior conjunction of the companion star), including the eclipse period. 
The jet precession phase in the observation is predicted based on the ephemeris of \citet{gies02} to be $\phi_{\rm jet} = 0.214$--$0.244$, where $\phi_{\rm jet} = 0$ corresponds to the epoch at which the approaching jet is inclined the most towards our line of sight. In this phase, the Doppler shifts of the approaching jet ($z_{\rm b}$) are expected to vary in the range of $-0.10 \lesssim z_{\rm b} \lesssim -0.09$ and that of the receding jet  ($z_{\rm r}$) in the range of $0.16 \lesssim z_{\rm r} \lesssim 0.17$, assuming the \citet{gies02} model, which considers the nodding motion with a $\sim6.3$-day period \citep{katz1982} in addition to the precession with a $\sim162$-day period.
However, 
the Doppler shifts $z_{\rm b}$ and $z_{\rm r}$ measured in our fits show a significant deviation from these predicted values ($-0.07 \lesssim z_{\rm b} \lesssim -0.04$ and $0.12 \lesssim z_{\rm r} \lesssim 0.14$; see below). 
This could be due to the uncertainties in the ephemeris enhanced by the large time gap between the observations by \citet{gies02} and ours, and the breaks in the precessional phase observed by \citet{cherepashchuk22}. The updated ephemeris obtained from this XRISM observation and a simultaneous optical spectroscopic campaign will be reported in a different paper \citep{sakai2025}.

We reduced the data following the XRISM Quick Start Guide v2.3\footnote{\url{https://heasarc.gsfc.nasa.gov/docs/xrism/analysis/quickstart/}}. We started with the cleaned event file provided by the XRISM team, processed with the pipeline processing version of 03.00.013.009, and used Heasoft version 6.34 along with the latest version of XRISM Calibration Database available as of 2024 September 30. In this study, we adopted only "High-resolution primary (Hp)" events for Resolve, which are best calibrated for energy. We removed the events recorded at pixel number 27, which was reported to show some unpredictable gain variation, as well as those at pixel number 12, which is the calibration pixel. The response matrix file (RMF) was generated with \texttt{rslmkrmf} with the file type option of  ``extra large'' ({\tt whichrmf = X}), based on the cleaned event file for the entire observation period. The ancillary response file (ARF) was generated with \texttt{xaarfgen}, assuming a point-like source at the aim point as input. 
To produce the spectrum, we adapted the data above 2 keV, below which X-ray signals are mostly absorbed by the closed gate valve. 
In this work, we did not subtract the non-Xray background (NXB), the cosmic X-ray background (CXB), nor the Galactic ridge X-ray emission (GRXE), which we found to be negligible. The NXB was estimated with {\tt rslnxbgen} to be $\lesssim 1$\% of the source flux over the 2--10 keV band. 
The contributions of the CXB and the GRXE were found to be one order of magnitude smaller than that of the NXB, assuming the CXB spectrum given in \citet{kushino2002} and the spatial profile and the spectrum of the GRXE provided by \citet{uchiyama2013}. 

For the Xtend data, source events were extracted from a circular region with a $1'$ radius centered on the source position. The background contribution was found to be negligibly small ($\lesssim 0.5$\% across the entire 0.5--10 keV range) and thus was not subtracted. 
Light curves were created to investigate flux variability including that below 2 keV, where the Resolve has limited sensitivity. However, spectral extraction was not performed, as the primary objective of this study was to resolve emission lines from the jets.

\begin{figure}
 \begin{center}
  \includegraphics[width=8.7cm]{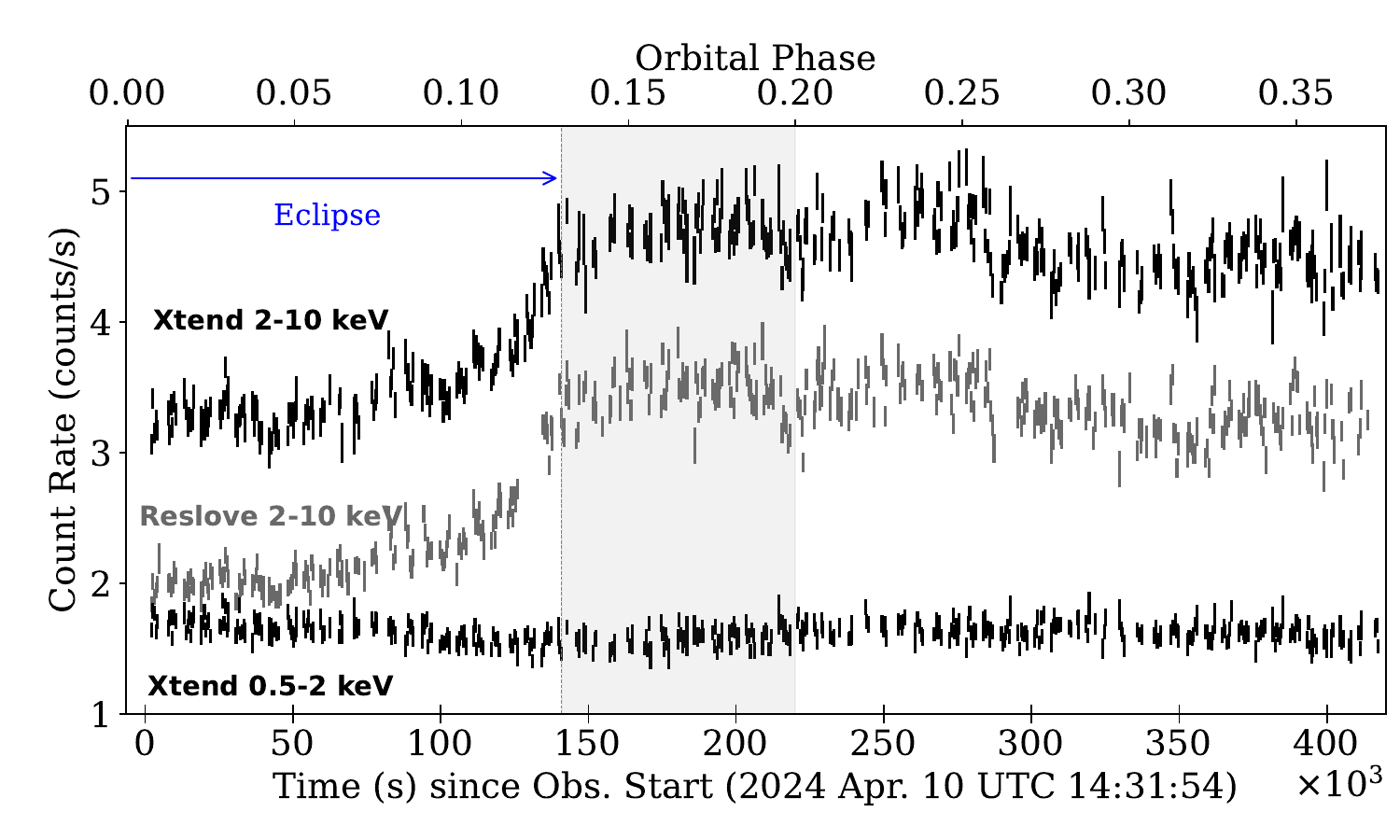} 
 \end{center}
\caption{Resolve (gray) and Xtend (black) light curves with 512 s bins. The errors indicate 1$\sigma$ statistical uncertainties. The shaded region indicates the time interval used for the spectral analysis described in Sec.~\ref{subsec:and_Fe_vs_Si}. 
{Alt text: a graph with error bars in the x- and y-axis directions. 
The x-axis shows the time and orbital phase and the y-axis shows the count rate. The top panel plots three different datasets.} 
}\label{fig:lc}
\end{figure}

\section{Analysis and Results}
\label{sec:ana}

Figure~\ref{fig:lc} presents the Resolve light curve in 2--10 keV and Xtend light curves in 0.5--2 keV and 2--10 keV over the entire observation period. 
The orbital phase derived by using the ephemeris in \citet{goranskij2011} is also indicated. In previous observations, a decrease of the X-ray flux due to the eclipse was observed in $|\phi_{\rm orb}| \lesssim $ 0.10--0.15 (e.g., \cite{kawai89, fillippova2006, cherepashchuk09,marshall2013, cherepashchuk2020}). In our observation, the source intensity initially stayed at a low level and gradually increased until $\phi_{\rm orb} \sim 0.13$ or $\sim 1.3 \times 10^5$~s from the start of observation. 
Considering this variation, we defined the first $\sim 1.3 \times 10^5$~s as the eclipse phase and the subsequent part as the non-eclipse phase. The 2--10 keV flux changed significantly as the eclipse ended, whereas the 0.5--2 keV flux remained almost constant over the entire observation period. This is consistent with the behavior observed during and outside eclipse in previous studies (e.g., \cite{marshall2013}). 

It should be noted, however, that the apparent constancy of the soft X-ray flux could be attributed not only to the intrinsic behavior of SS~433 (i.e., only the innermost, hottest part of the jet is obscured during the eclipse; e.g., \cite{kawai89}) but might also reflect some 
contribution from interstellar dust scattering. Given the high interstellar column: $N_{\rm H} = (1-2) \times 10^{22}$ cm$^{-2}$ (e.g., \cite{lopez2006,kubota10,marshall2013}), the optical depth for dust scattering can reach $\gtrsim$ 2 below 1 keV, assuming the scattering cross-section in \citet{drain2003}. 
Depending on the distance to the interstellar dust, the scattered-in component could introduce a time lag, which may affect the flux variation in the soft X-ray band, although in previous Chandra observations the dust halo is almost always a pure scattering loss without time-delayed soft components \citep{constantini22,corrales16}. Even in the eclipsing binary AX J1745.6$-$2901, where the absorption column density is an order of magnitude higher, the flux decrease in eclipses are still visible \citep{jin18}, so the effect may not play a main role.

\begin{figure}
 \begin{center}  
 \includegraphics[width=8.9cm]{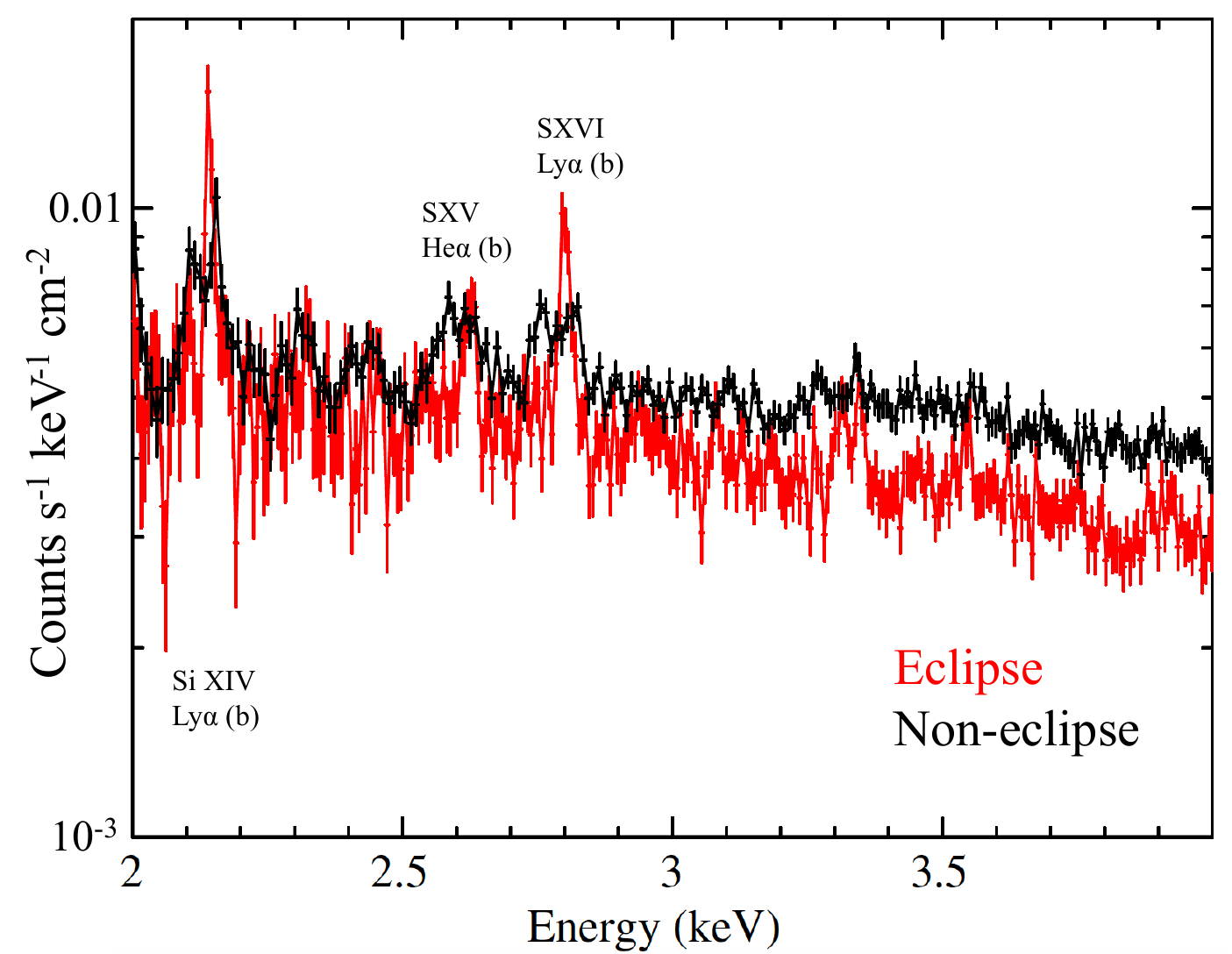} 
  \includegraphics[width=8.9cm]{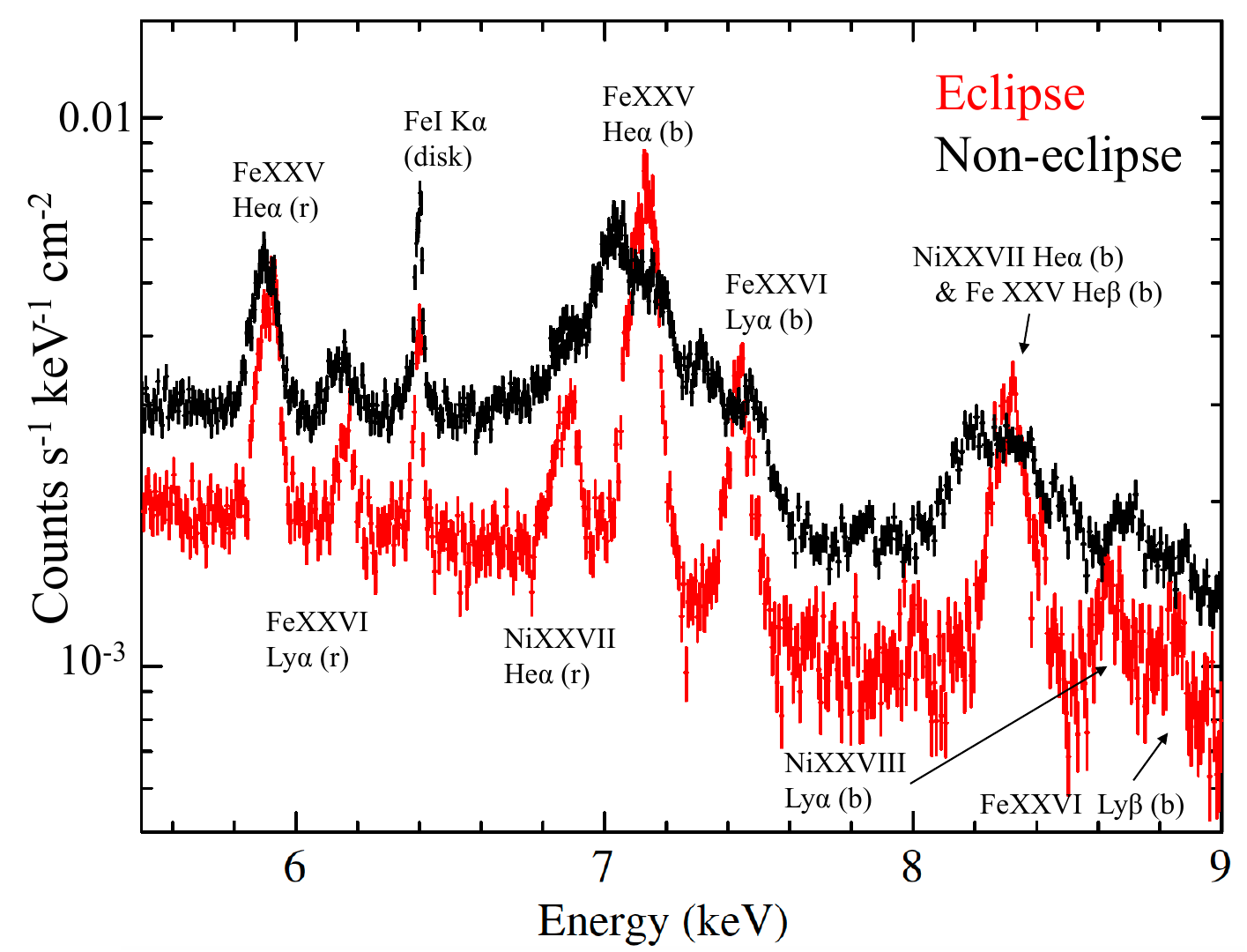} 
 \end{center}
\caption{Time-averaged Resolve 2--4 keV (top) and 5.5--9 keV (bottom) spectra, corrected for the effective area of the instrument, in the eclipse and non-eclipse phases. The identifications of main lines are also presented, where (r), (b), and (disk) denote those from the approaching jet, the receding jets, and the accretion disk (or the disk wind), respectively. Note that the non-eclipse phase is longer than the eclipse phase and therefore the lines in the former appear broader, likely due to variations in Doppler shifts caused by the jet precession. {Alt text: Two graphs with error bars in the x- and y-axis directions. Two data (in eclipse and non-eclipse phases) are presented in each graph. The x- and y-axes show the energy in units of keV and the count rate, respectively.} 
}\label{fig:spec}
\end{figure}

\begin{figure*}
 \begin{center}
  \includegraphics[width=8.8cm]{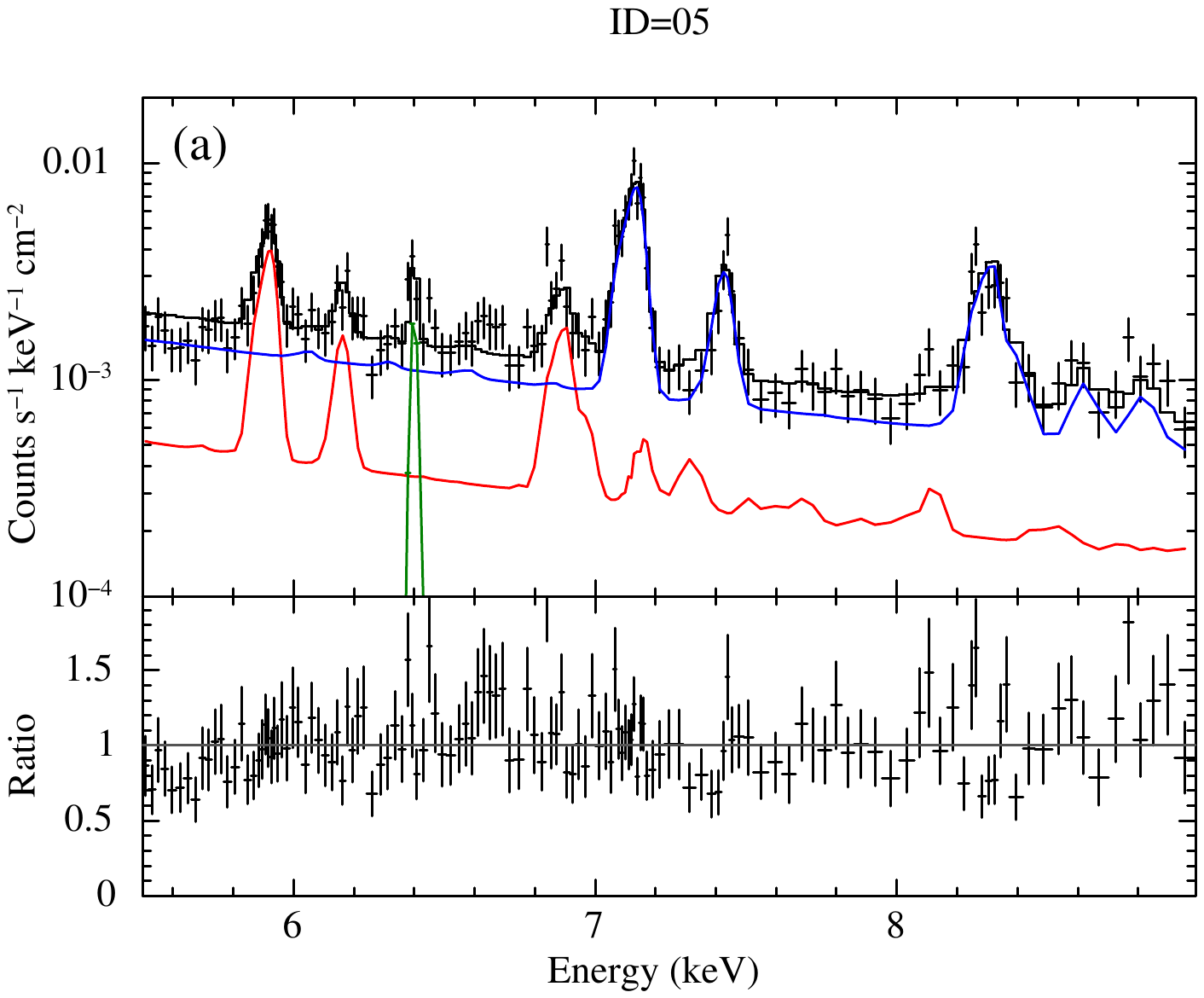} 
  \includegraphics[width=8.8cm]{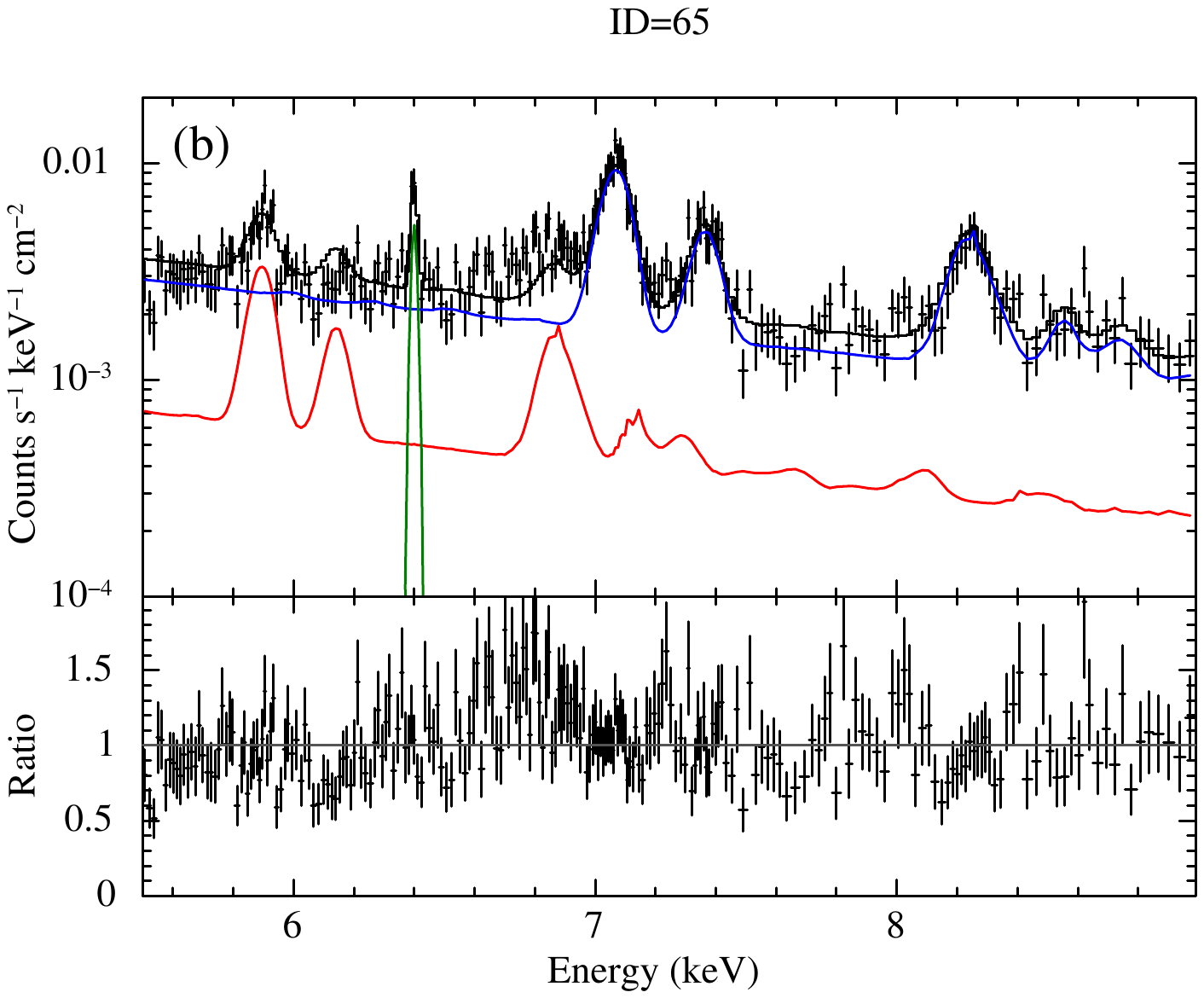} 
 \end{center}
\caption{Representative spectra in the (a) eclipse and (b) non-eclipse phases and their best-fit models. Contributions of the approaching jet, receding jets, and the narrow FeI K$\alpha$ line
are presented with the blue, red, and green lines, respectively. The spectra are corrected for the effective area of the instrument.
The data versus model ratios are shown in the bottom panels. 
{Alt text: Two figures lined-up horizontally, each with two graphs combined vertically. The upper panels show graphs with error bars in the x- and y-axis directions. The x and y axes show the time and the count rate, respectively. In the lower panels the y-axis is the data versus model ratio.} 
}\label{fig:spec_fit}
\end{figure*}

In Figure~\ref{fig:spec} we plot the time-averaged Resolve spectra in 2--4 keV and 5.5--9 keV obtained in the eclipse and non-eclipse phases. Doppler-shifted ionized Fe and Ni lines originating from the approaching and receding jets are present in both spectra, as well as a narrow FeI K$\alpha$ line, which is considered to be produced in the outer region of the accretion disk (\cite{kotani1996}, \cite{takagi2025}). The jet emission lines in the non-eclipse phase are significantly broader than those in the eclipse phase, likely due to the change in the Doppler shift by the jet precession. Because the non-eclipse phase has a longer duration than the eclipse phase, the lines should be more broadened. 
Remarkably, as clearly demonstrated by the eclipse-phase spectrum, which is less affected by Doppler shift variations than the non-eclipse phase spectrum, we successfully resolved emission lines even above $\sim 8$ keV, including the Ni XXVIII Ly$\alpha$ line and the Fe XXVI Ly$\beta$ line of the approaching jets. This result highlights the excellent energy resolution of Resolve as well as its high sensitivity at high energies.

\subsection{Time Variation of the Velocity Dispersion in the Fe K Band}
\label{subsec:ana_Fe_var}
To mitigate the line broadening effect by the jet precession, we divided the Resolve data into each continuous exposure and used the time-averaged spectra of the individual data segment for spectral analysis. We got 95 segments (ID$=$01--95 in the chronological order) in total and the typical net exposure per segment is $\sim$ 3500 s. Figure~\ref{fig:spec_fit} shows representative spectra in the eclipse and non-eclipse phases. 
We performed spectral modelling of these spectra on XSPEC version 12.14.1d \citep{arnaud96}. In this analysis, we excluded 9 segments 
(IDs$=$ 17, 19, 36, 38, 50, 57, 59, 76, and 78) 
which have an exposure of $< 500$ s and have too low statistics. We grouped the spectra to contain at least 1 count in each bin and adopted an energy range of 5.5--9 keV to focus on the Fe K band. We created the RMF and ARF files for the individual data segments in the same manner as those for the entire period of the observation (see Section~\ref{sec:obs}). 

We adopted the {\tt bvapec} model, the optically-thin collisionally ionized plasma model APEC \citep{smith01} with variable abundances and additional Gaussian line broadening, to account for the jet emission. The input parameters of {\tt bvapec} are the plasma temperature $kT$, the abundances of atoms, the Doppler shift $z$, the 1$\sigma$ width ($\sigma_{\rm jet}$) of the additional line broadening, and the normalization. We allowed the Ni abundance $A_{\rm Ni}$ to vary, as it has been reported to exceed the solar values in previous studies (e.g., \cite{marshall2002,brinkmann2005,marshall2013,medvedev2018}). For the other atoms, the solar abundance was assumed. Here, we adopted the solar abundance table given by \citet{lodders09} and the cross-section table by \citet{verner96}.  
We combined two {\tt bvapec} components to reproduce the emissions from the approaching and receding jets (hereafter {\tt bvapec\_b} and {\tt bvapec\_r}, respectively). 
All the parameters except for the normalization and $z$ were linked between the two jet components. We tested spectral fitting of several spectra with these parameters unlinked, but no significant differences were detected between the two jet components. The $kT$ and $\sigma_{\rm jet}$ values of the receding jet typically had large 90\% uncertainties of $\sim \pm 2$ keV and $\sim \pm$ 1000 km s$^{-1}$, respectively, due to its small contribution to the total flux and the limited data statistics. The Doppler shifts of the two jet components were constrained with nearly the same precision as when $kT$ and $\sigma_{\rm jet}$ were linked, and remained consistent within their 90\% errors.
A narrow Gaussian component was also added to account for the FeI K$\alpha$ line. We fixed the line center energy and the 1$\sigma$ line width at 6.4 keV and 10 eV, respectively. This line width was determined by fitting the line in the time-averaged spectrum for the entire non-eclipse phase with a Gaussian model. 
The spectral model that we used in our analysis is thus expressed as {\tt bvapec\_r+bvapec\_b+gauss}. 

\begin{table}
  \tbl{Best-fit parameters of the representative spectra in the eclipse and non-eclipse phases.\footnotemark[$*$] }{%
  \begin{tabular}{cccc}
      \hline
     & & Eclipse (ID$=$05) & Non-eclipse (ID$=$65)  \\  \hline
      \multicolumn{2}{c}{Start\footnotemark[$\dag$] (s)}  & 22884 & 281841  \\ 
      \multicolumn{2}{c}{End\footnotemark[$\dag$] (s)}  & 26376 &  285375 \\ 
      \hline
    Component & Parameter &   \multicolumn{2}{c}{Value} \\  \hline
      {\tt bvapec\_b} 
           & $kT$ (keV) & $6.3 \pm 0.4$ & $7.2^{+0.4}_{-0.3}$ \\
           & $A_{\rm Ni}$ & $8^{+2}_{-1}$ & $8 \pm 1$  \\
           & $z$  & $-0.0620 \pm 0.0005$ & $-0.0544^{+0.0005}_{-0.0004}$ \\
& $\sigma_{\rm jet}$\footnotemark[$\ddag$] (km s$^{-1}$) & $(1.0 \pm 0.1) \times 10^3$ & $(1.8 \pm 0.2) \times 10^3$ \\
& norm. & $0.091 \pm 0.006$ & $0.163 \pm 0.009$ \\
       {\tt bvapec\_r} & $z$ & $0.1303^{+0.0007}_{-0.0008}$ & $0.134 \pm 0.001$ \\
           & norm. & $(4.7 \pm 0.6) \times 10^{-2}$ & $(5.9 \pm 0.9) \times 10^{-2}$ \\
     {\tt gauss} & $E_{\rm cen}$ (keV)& 6.4 (fix) & 6.4 (fix) \\
                 & $\sigma$ (eV) & 10 (fix) & 10 (fix) \\
                 & norm. $(10^{-5})$ & $6^{+3}_{-2}$ & $14^{+4}_{-3}$ \\
      \hline
                 \multicolumn{2}{c}{C-stat/d.o.f.} & 1907/2334 & 2788/3507 \\
      \hline
    \end{tabular}}\label{tab:param}
\begin{tabnote}
\footnotemark[$*$] Model: {\tt bvapec\_r+bvapec\_b+gauss}, where b and r denote the approaching and receding jets, respectively. \\ 
\footnotemark[$\dag$] Start and end times of the exposure. Seconds from the start of the XRISM SS 433 observation: 2024 Apr. 10 UTC 14:31:54.  \\
\footnotemark[$\ddag$]  1$\sigma$ width of Gaussian line broadening. 
\end{tabnote}
\end{table}

\begin{figure}
 \begin{center}
  \includegraphics[width=9.cm]{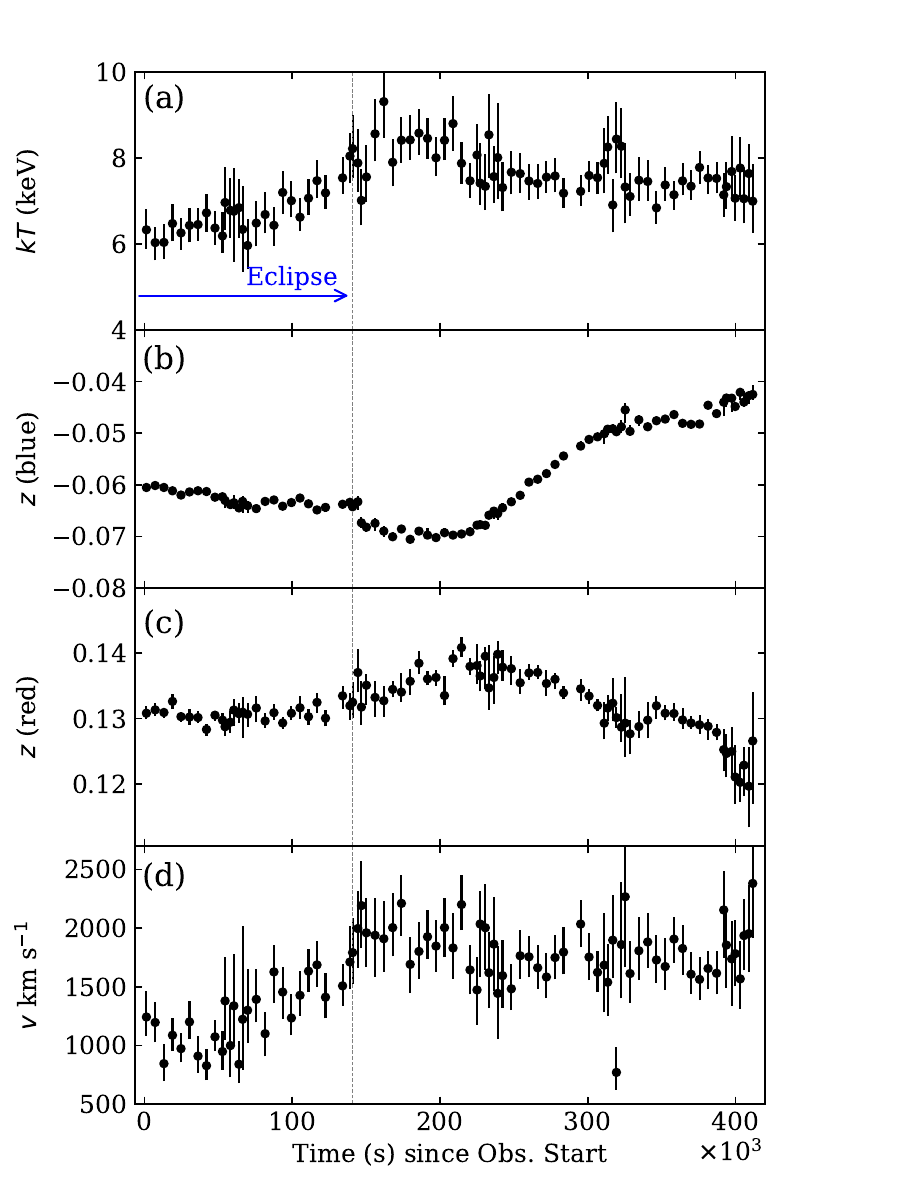} 
 \end{center}
\caption{Evolution of the jet parameters: the plasma temperature (a), the Doppler shifts of approaching and receding jet components (b and c, respectively), and their velocity dispersion (d).
{Alt text: four graphs with error bars in the x- and y-axis directions are aligned vertically. The x-axis shows the time and y-axis shows the jet parameters. }}
\label{fig:sigma}
\end{figure}

Figure~\ref{fig:spec_fit} shows the best-fit models for the the two representative spectra in the eclipse and non-eclipse phases, and Table~\ref{tab:param} gives their best-fit parameters. 
As seen in previous studies (e.g., \cite{kawai89, fillippova2006, kubota10b}) and as expected from the XRISM light curves, $kT$ increased from the eclipse phase to the non-eclipse phase. 
We found that the jet emission lines in the non-eclipse phase were significantly broader than those in the eclipse phase. The velocity dispersion ($\sigma_{\rm jet}$) was estimated to be $\sim 1000$ km s$^{-1}$ from the spectrum in the eclipse phase and $\sim$ 1700 km s$^{-1}$ from that in the non-eclipse phase. In Figure~\ref{fig:sigma}, we present the time variation of the main parameters characterizing the jet components over the entire observation. Since the Ni abundance did not change significantly during the observation, we omitted from in the figure. The velocity dispersion gradually increased during the egress ($\sim 5\times 10^4$ s to $\sim 1.2 \times 10^5$ s), where the flux increased gradually (Fig.~\ref{fig:lc}), and became almost constant during the non-eclipse phase. We estimated the weighted averages of $\sigma_{\rm jet}$ over the eclipse phase before the start of egress (which we defined as $7 \times 10^4$ s) and over the non-eclipse phase to be 
$1020 \pm 40$ km s$^{-1}$ and $1740 \pm 30$ km s$^{-1}$, 
respectively, where the errors represent the weighted standard deviations. 

Although the model was able to reproduce the overall profile of the individual spectra and gave acceptable fit, small residual structures often remained around 6.6--6.8 keV, particularly in the non-eclipse spectra. A similar structure was reported in previous Chandra observation \citep{medvedev2019}, where it was modeled with a broad Gaussian component. Motivated by this work, we added a Gaussian component at $\sim 6.7$ keV. We found that including the Gaussian reduced the C-statistic by $\lesssim 50$, while the best-fit parameters remained consistent with those obtained without the Gaussian component within their 90\% confidence ranges, and the increasing trend of $\sigma_{\rm jet}$ was unchanged.

\subsection{Comparison of the Velocity Dispersions in the Fe K and Si/S K Bands}
\label{subsec:and_Fe_vs_Si}

We also assessed the variation of the velocity dispersion with a different approach: by comparing the lines around the Fe K band (in 5.5--9 keV) with those around the Si/S K band (in 2--4 keV), which was adopted in the previous work by \citet{namiki2003}. As noted in the previous section, longer time integrations can make the line broadening due to Doppler shifts from jet precession and nodding motion non-negligible. Nevertheless, to demonstrate the quality of the Resolve data with a moderate integration time in both the Fe and Si K line regions,  
and to enable direct comparison with previous results including \citet{namiki2003}, we present the analysis of the time-averaged spectrum. 

To achieve sufficient statistics but avoid the jet obscuration due to the eclipse and minimize line broadening by the precession, we adopted the data in $\phi_{\rm orb} = 0.13$--$0.20$, whose net exposure is $\sim 40$ ks. In this period, the variation in the 
Doppler shifts of the approaching and receding jets 
was relatively small ($\Delta z_{\rm b} \sim -0.006$ and $\Delta z_{\rm r} \sim 0.008$, estimated from the time-resolved spectral analysis in Sec.~\ref{subsec:ana_Fe_var}) compared with the other periods. 
We adopted the same spectral model as the time-resolved analysis, but multiplied the neutral absorption model {\tt TBabs} to consider the interstellar absorption, which can affect the spectral profile mainly at low energies. The tables for the elemental abundances and cross sections adopted for {\tt tbabs} were the same as those for {\tt apec}. 
The model was applied to the spectrum in the 2--4 keV band and in the 5.5--9 keV band independently.

First, we describe the analysis and results of the 5.5--9 keV band. 
We varied $kT$ and $\sigma_{\rm jet}$ of the two {\tt vapec} components independently to investigate whether there are differences in plasma properties of the two jets. We found that it is difficult to constrain $N_{\rm H}$ of {\tt TBabs}, which mainly affects the spectral profile at lower energies. Previous Chandra observations reported $N_{\rm H} = (1$--$3) \times 10^{22}$ cm$^{-2}$ \citep{marshall2002,lopez2006,marshall2013}. We fixed $N_{\rm H}$ at $2.07 \times 10^{22}$ cm$^{-2}$ for the both energy bands, which was determined by Chandra data \citep{lopez2006}. We also tested different fixed $N_{\rm H}$ values from $0$ to $5 \times 10^{22}$ cm$^{-2}$ but found that the best-fit values of $kT$, $\sigma_{\rm jet}$, $A_{\rm Ni}$, $z_{\rm r}$, and $z_{\rm b}$ exhibited only negligible changes, much smaller than their 90\% uncertainties. 

Although this model roughly reproduced the overall spectral profile in 5.5--9 keV, significant residuals were found to remain around 6.2--7.0 keV, similar to those seen in the time-averaged spectra. To account for these structures, we attempted to include an  additional {\tt bvapec} component and also tested multi-temperature plasma emission models. However, because these features are localized within a limited energy range, it was difficult to reproduce them with additional jet emission components, and the fit was not improved. 
Instead, following the approach used for the time-averaged spectra, we added two Gaussian components at $\sim$ 6.4 keV and $\sim$ 6.9 keV, which led to a substantial improvement in the fit quality.

\begin{figure}
 \begin{center}
   \includegraphics[width=8.8cm]{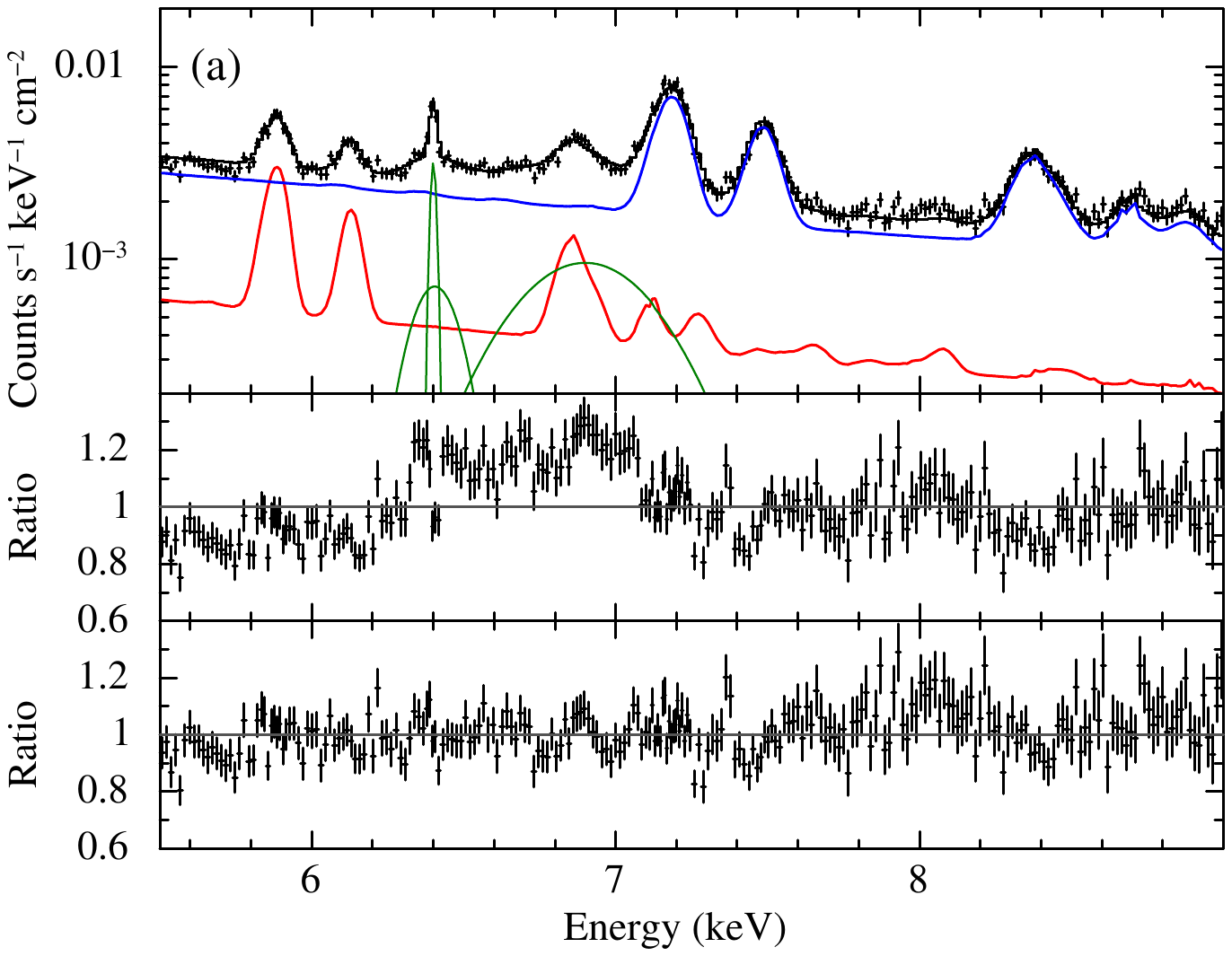} 
  \includegraphics[width=8.8cm]{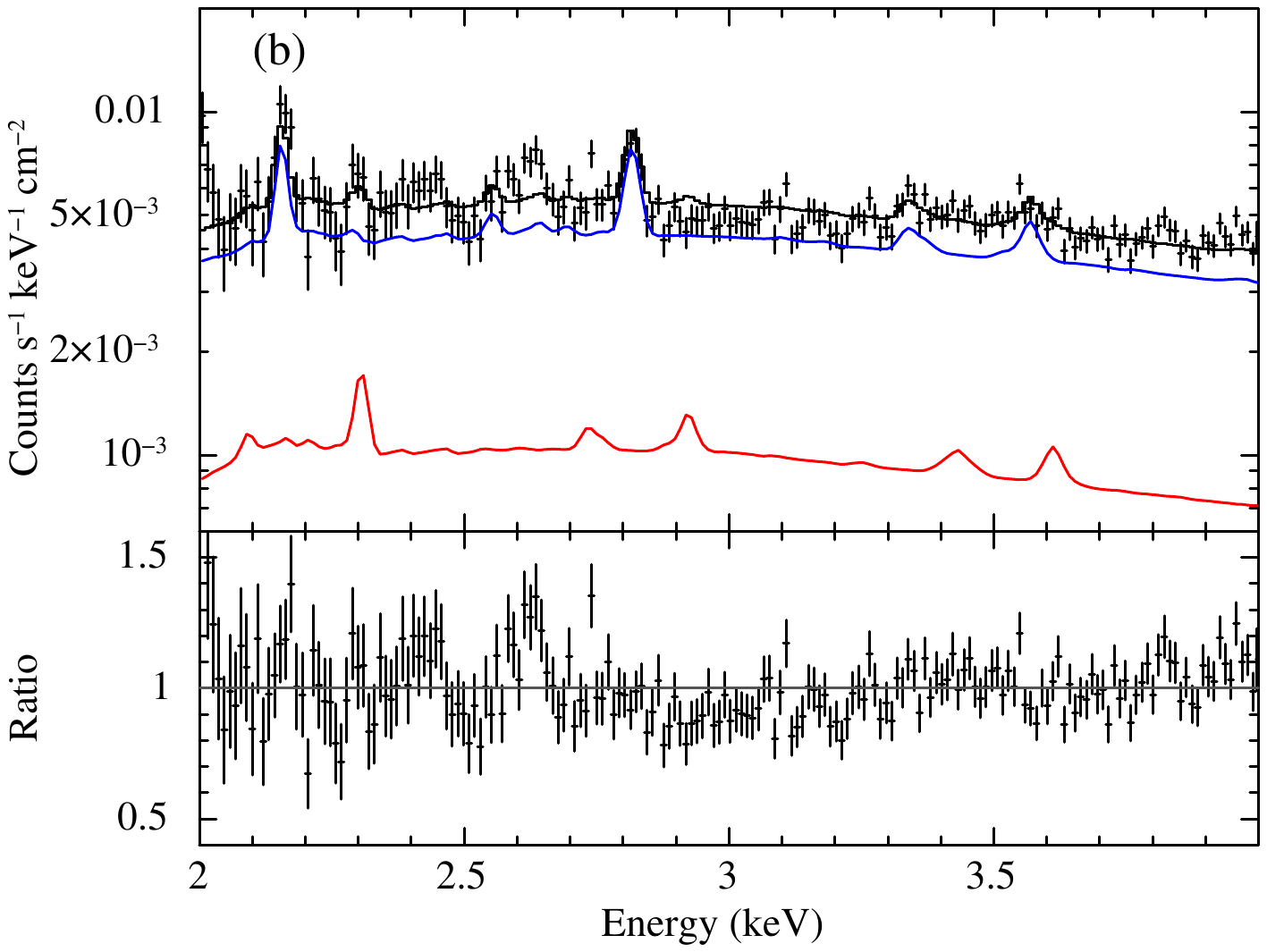} 
 \end{center}
\caption{Time-averaged spectra for $\phi_{\rm orb} = 0.13$--$0.20$ in (a) 5.5--9 keV and (b) 2--4 keV and the best-fit models, corrected for the effective area of the instrument. The contributions of the approaching jet, receding jet, and the Gaussian line components 
are shown with the blue, red, and green lines, respectively. The middle and bottom panels in (a) 
present the data versus model ratio of the best-fit models without and with two broad Gaussian components, respectively.
{Alt text: two figures lined-up vertically, each with two graphs combined vertically. The upper panels show graphs with error bars in the x- and y-axis directions. The x and y axes show the time and the count rate, respectively. In the lower panels the y-axis is the data versus model ratio.}
}\label{fig:Si_and_Fe_fit}
\end{figure}

The data and the best-fit model are shown in Figure~\ref{fig:Si_and_Fe_fit}(a) and the best-fit parameters are listed in Table~\ref{tab:Si_and_Fe_fit}, respectively. The residuals and parameters for the model without the broad Gaussian components are also presented in these figure and table. 
We detected a significant difference in $kT$ and $\sigma_{\rm jet}$ between the two jet components. 
The $kT$ value of the receding jet was lower than that of the approaching jet by $0.7 \pm 0.6$ keV, and the $\sigma_{\rm jet}$ value for the receding jet ($\sigma_{\rm jet(r)}$) was smaller by $500 \pm 200$ km s$^{-1}$ compared to that for the approaching jet ($\sigma_{\rm jet(b)}$).
Note that this difference in $\sigma_{\rm jet}$ is unlikely to be fully explained by 
differences in Doppler-shift variability between the two jets. In fact, the variation of the measured $z_{\rm b}$ value in Fig.~\ref{fig:sigma} is slightly smaller than that of $z_{\rm r}$ by $\sim 0.002$ (or $\sim$ 600 km s$^{-1}$), which is in the opposite direction to the observed behavior of $\sigma_{\rm jet}$ in our spectral fits.

Next, we analyzed the 2--4 keV spectrum using the same model as for the 5.5--9 keV spectrum, excluding the narrow and broad Gaussian components. We first allowed 
$A_{\rm Ni}$, $kT$, $\sigma_{\rm jet(r,b)}$, $z_{\rm r,b}$, and the normalization of the {\tt bvapec} components to vary freely, in the same manner as the analysis of the 5.5--9 keV data. We also allowed $N_{\rm H}$ of {\tt TBabs} to vary. As a result, $kT$ of the receding jet became an anomalous value of $>60$ keV, which has never been obtained in previous studies, and the normalization of the receding jet became $\sim 3$ times larger than that of the approaching jet. The situation did not change when we fixed other parameters such as $N_{\rm H}$. We therefore linked the $kT$ values of the approaching and receding jet components. In this model, a very small normalization of the receding jet component ($\sim 10^{-6}$) was favored. This resulted in a flux ratio of the receding to approaching jet that is smaller by several orders of magnitude compared to the value derived from the 5.5--9 keV data, and inconsistent with previous studies using Chandra (e.g., \cite{marshall2013}). In these trials, $\sigma_{\rm jet(r)}$ and $z_{\rm r}$ were not constrained at all.

In summary, the parameters of the receding jet component could not be constrained with sufficient accuracy, most likely due to the limited statistics and its minor contribution to the total emission. Consequently, we adopted the flux ratio determined from the normalization parameters of the approaching and receding jets in the 5.5--9 keV spectral fit. 
Constraining $A_{\rm Ni}$ was also found to be difficult, because no strong lines are present in 2--4 keV. Indeed, allowing $A_{\rm Ni}$ to vary yielded only a very weak constraint was obtained, $A_{\rm Ni} < 60$ in the above results. We thus assumed $A_{\rm Ni}$ at 6.3, determined from the 5.5--9 keV data. 
We adopted $z_{\rm r} = 0.136$, which was determined from the 5.5--9 keV data, and linked $\sigma_{\rm jet(r)}$ to $\sigma_{\rm jet(b)}$. Note that previous Chandra observations have demonstrated that the Doppler shifts of both low-energy and high-energy emission lines are in agreement \citep{marshall2002,marshall2013}. We confirmed that fixing $z_{\rm r}$ at different values in 0.1--0.2 and $A_{\rm Ni}$ in 1--20, or unlinking $\sigma_{\rm jet(r)}$ from  $\sigma_{\rm jet(b)}$ does not change the resultant best-fit parameters, including $\sigma_{\rm jet(b)}$.

\begin{table*}
      \tbl{Best-fit parameters of the time-averaged spectra during $\phi_{\rm orb} = 0.13$--$0.20$.\footnotemark[$*$]}{%
  \begin{tabular}{ccccc}
      \hline
   Component & Parameter &   \multicolumn{3}{c}{Value} \\  
   &  & 2--4 keV & 5.5--9 keV & 5.5--9 keV \\ \hline
     {\tt TBabs} 
           & $N_{\rm H}$ (10$^{22}$ cm$^{-2}$) & $2.7^{+0.3}_{-0.6}$  & 2.07 (fixed) & 2.07 (fixed)  \\
       {\tt bvapec\_b} 
           & $kT$ (keV) & $5.1^{+2.0}_{-0.6}$ & $8.2 \pm 0.2$ & $8.4 \pm 0.2$  \\
           & $A_{\rm Ni}$ & 6.3 (fixed) & $6.8 \pm 0.4$ & $6.3^{+0.5}_{-0.4}$  \\
           & $z$ & $-0.0693 \pm 0.0009$ & $-0.0691 \pm 0.0002$ & $-0.0695^{+0.0002}_{-0.0003} $ \\
    & $\sigma_{\rm jet(b)}$ (km s$^{-1}$) & $1.3^{+0.3}_{-0.4} \times 10^3$ & $(2.00 \pm 0.05) \times 10^3$& $(1.90 \pm 0.08) \times 10^3$ \\
           & norm. & $0.13^{+0.01}_{-0.02}$ & $0.156^{+0.002}_{-0.004}$ & $0.154 \pm 0.005$ \\
       {\tt bvapec\_r} 
           & $kT$ (keV) & -\footnotemark[$\ddag$] & $8.2 \pm 0.3$ & $7.7 \pm 0.6$ \\
           & $z$ & 0.136 (fixed) & $0.1363^{+0.0004}_{-0.0005}$ & $0.136 \pm 0.005$\\
               & $\sigma_{\rm jet(r)}$ (km s$^{-1}$)\footnotemark[$\dag$] & -\footnotemark[$\ddag$] & $(1.7 \pm 0.2) \times 10^3$ & $(1.4 \pm 0.2) \times 10^3$ \\
           & norm. ($10^{-2}$) & -\footnotemark[$\S$] & $6.5 \pm 0.4$ & $5.2 \pm 0.4$ \\
     {\tt gauss} & $E_{\rm cen}$ (keV)& - & 6.4 (fix) & 6.4 (fix) \\
                 & $\sigma$ (eV) & - & 10 (fix) & 10 (fix) \\
                 & norm. $(10^{-4})$ & - & $1.13 \pm 0.01$ & $0.86 \pm 0.01$ \\
     {\tt gauss} & $E_{\rm cen}$ (keV)& - & - & $6.4^{+0.03}_{-0.02}$\\
                 & $\sigma$ (eV) & - & - & $80^{+40}_{-30}$ \\
                 & norm. $(10^{-4})$ & - & - & $1.5^{+0.5}_{-0.4}$ \\
     {\tt gauss} & $E_{\rm cen}$ (keV)& - & - &$6.90 \pm 0.03$ \\
                 & $\sigma$ (eV) & - & - & $220 \pm 40$ \\
                 & norm. $(10^{-4})$ & - & - & $5.6 \pm 0.8$ \\
      \hline
                 \multicolumn{2}{c}{C-stat/d.o.f.} & 4319/3994 & 8029/6989 & 7422/6983 \\
      \hline
    \end{tabular}}\label{tab:Si_and_Fe_fit}
\begin{tabnote}
\footnotemark[$\dag$]  1$\sigma$ width of the Gaussian line broadening. \\
\footnotemark[$*$] Model: {\tt TBabs*(bvapec\_r+bvapec\_b + gauss + gauss + gauss)}. The Gaussian components were omitted for 2--4 keV spectrum. For the 5.5--9 keV data, best-fit parameters with and without the two broad gaussian components are listed in the left and right columns, respectively.\\ 
\footnotemark[$\ddag$]  Linked to the values of the approaching jet ({\tt bvapec\_b}). \\ 
\footnotemark[$\S$]  Assumed to be 0.34 times the normalization of {\tt bvapec\_b} (see text). \\ 
\end{tabnote}
\end{table*}

\begin{figure}
 \begin{center}
  \includegraphics[width=8.8cm]{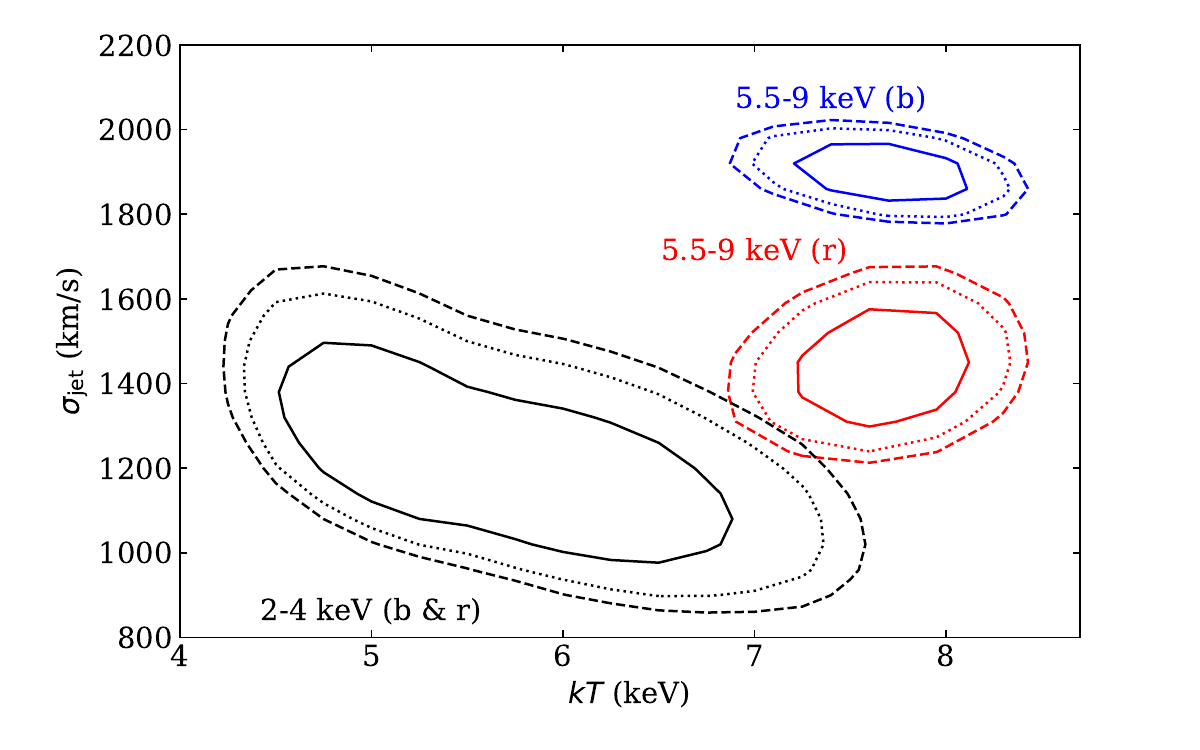} 
 \end{center}
\caption{Confidence contours for the plasma temperature $kT$ and the velocity dispersion $\sigma_{\rm jet}$ obtained in the 2--4 keV band (black) and those obtained in 5.5--9 keV for the approaching jet (blue) and the receding jet (red) components. Solid, dotted, and dashed lines show the 68\%, 90\%, and 95\% confidence levels, respectively. 
Note that in the 2–4 keV band, $\sigma_{\rm jet}$ for both jet components was linked (see text), and its value was primarily constrained by the emission lines from the dominant approaching jet component (see Fig.~\ref{fig:Si_and_Fe_fit}). 
Alt text: A plot showing confidence contours for three different parameter sets in the $kT$ (plasma temperature; horizontal axis) and $\sigma_{\rm jet}$ (velocity dispersion; vertical axis) plane.
}\label{fig:cont}
\end{figure}

The 2--4 keV data and the best-fit model are shown in Figure~\ref{fig:Si_and_Fe_fit}(b) and the best-fit parameters are listed in Table~\ref{tab:Si_and_Fe_fit}, respectively. 
The plasma temperature estimated in the 2--4 keV band was $kT=5.1^{+2.0}_{-0.6}$ keV, which was somewhat smaller than in the 5.5--9 keV region ($8.4 \pm 0.2$ keV and $7.7 \pm 0.6$ keV for the approaching and receding jets, respectively), suggesting that the Si and S lines were produced in cooler plasma than Fe and Ni lines. The $\sigma_{\rm jet}$ value in the 2--4 keV spectrum ($1300^{+300}_{-400} $ km s$^{-1}$) was significantly smaller than that of the approaching jet measured from the 5.5--9 keV spectrum ($1900 \pm 80$ km s$^{-1}$) but comparable with that of the receding jet from 5.5--9 keV ($1400 \pm 200$ km s$^{-1}$). 
Figure~\ref{fig:cont} displays the confidence contours in the plane of $\sigma_{\rm jet}$ and the plasma temperature $kT$, which mainly determines the strengths and profiles of the individual emission lines. Significant differences in the $\sigma_{\rm jet}$ values between the approaching and receding jets in 5.5--9 keV and between the approaching jet in 5.5--9 keV and in 2--4 keV are confirmed in this plot. 

In the above analysis, we used the solar abundance table by \citet{lodders09} following other recent XRISM/Resolve studies (e.g., \cite{xrism2025_centaurus}), whereas previous studies using the {\tt TBabs} model often adopted the table by \citet{wilms2000}. When we repeated the fit using the \citet{wilms2000} abundance table, the derived $N_{\rm H}$ measured in the 2--4 keV band and $A_{\rm Ni}$ in the 5.5--9 keV band became larger than those obtained with the \citet{lodders09} table ($N_{\rm H} = 3.8 \pm 0.4 \times 10^{22}$ cm$^{-2}$ and $A_{\rm Ni}=9.4^{+0.8}_{-0.6}$, respectively), 
while the other parameters remained consistent within the 90\% confidence range. The difference in $A_{\rm Ni}$ is likely due to the Ni/Fe abundance ratio given in \citet{wilms2000} is $\sim 1.5$ times smaller than that in \citet{lodders09}.

\section{Discussion and Conclusion}
\label{sec:discussion}
Thanks to the excellent energy resolution and good effective area of the XRISM/Resolve, we have successfully resolved Doppler-shifted emission lines from the jets, including the Fe XXVI Ly$\alpha$ line, which is less affected by line blending. From the time-resolved spectroscopy around the Fe K band, the jet emission lines in the non-eclipse phase were found to be significantly broader than those in the eclipse phase. 
A clear increasing trend in velocity dispersion was 
observed in the last part of the eclipse (egress) phase when the X-ray flux gradually increased. 
Given that the hottest part (closest to the compact object) of the X-ray emitting region in the jet is hidden by the companion star during the eclipse (e.g., \cite{kawai89}), the observed variation of the line width indicates that the hotter region of the jet (closer to the jet base) has a larger velocity dispersion than the colder region (farther from the jet base).
This is consistent with what \citet{namiki2003} suggested using a Chandra spectrum. 

The observed variation of the velocity dispersion is significant despite possible systematic uncertainties in its measurement. The line could be broadened due to the change in the line-of-sight velocity by the jet precession. However, it is estimated to be $100$--$200$ km s$^{-1}$ in each data segment (with the duration of $\sim 3500$ s), according to the previous studies of the jet precession motion (\cite{katz1982, gies02}). This is much smaller than the observed variation of velocity dispersion.  
The binary motion of the compact object was estimated to be a few hundred km s$^{-1}$ in our line-of-sight direction (e.g., \cite{kubota10}), and its change in the $\sim$ 5-day observation of XRISM should be less than $\sim 100$ km s$^{-1}$, which is significantly smaller than the observed variation of the velocity dispersion. 
In addition, the changes in the line shifts due to the motions of the satellite and the Earth are within of order 1--10 km s$^{-1}$ and therefore are negligible.

We have also confirmed the spatial variation of the velocity dispersion along the jet axis by adopting the same approach as \citet{namiki2003}: comparison of the line widths of emission lines at high and low energies, using a time-averaged spectrum in the non-eclipse period. 
In this analysis, the line widths of the approaching and receding jet components was constrained separately in the Fe and Ni line region. By contrast, in the Si and S region, the receding jet component remained largely unconstrained and only the width of the approaching jet was determined. The ionized Fe and Ni K lines were found to be significantly broader than the Si and S K lines (Sec.~\ref{subsec:and_Fe_vs_Si}), when comparing the approaching jet components. 
The velocity dispersions of the Si and S K line regions and the Fe and Ni K line region were estimated to be $\sim 1300$ km s$^{-1}$ and $\sim 1900$ km s$^{-1}$, respectively, which are consistent with the values in \citet{namiki2003}. 

Possible mechanisms for the observed line broadening includes the thermal Doppler broadening as well as variations in the Doppler shifts caused by jet precession and nodding motion. Based on the plasma temperatures determined from the spectral analysis, the thermal broadening for Si and Fe is estimated to be less than 100 km s$^{-1}$. The effects of jet precession and nodding motion can be inferred from the time-resolved analysis in 5.5--9 keV, yielding $\Delta z \sim 0.006$ for the approaching jet, corresponding to half widths of $\sim$ 900 km s$^{-1}$. 
This value is slightly smaller than, or at least comparable to, the line widths measured in the Si K band. 
In contrast, the approaching jet component in the Fe K band exhibits significantly larger line widths, suggesting the presence of additional velocity broadening mechanisms beyond the thermal broadening and the jet precession and nodding motion. 
Given that the Si emission region are thought to be located farther out than the Fe emission region, the above result can be attributed to a decrease of a velocity dispersion arising from processes other than thermal motions or the precession/nodding motion of the jet, along the direction of the jet flow.

Note that the Doppler shift variation was measured only from the time-resolved analysis of the Fe K region, and it might be possible that the amplitude of the variation could be different in the Si K region, due to the propagation delay and intrinsic change of the jet precession and nodding motion. Considering that the X-rays are emitted within $10^{12}$ cm \citep{kotani1996}, the propagation time of jet plasma between the Fe K and Si K line-emitting regions is estimated to be $\lesssim 100$ s using the jet speed of 0.26$c$, which is far shorter than the integrated time of the time-averaged spectrum analyzed in Section~\ref{subsec:and_Fe_vs_Si}. Consequently, any variations in the jet precession and nodding motion that occur in the Fe K region should propagate to the Si K region. Therefore, the Fe K region and Si K region should have the same amplitude of the Doppler shift variation, unless these motions are systematically changed in the Si K region from the Fe K region. Such change of jet motions, however, has not reported to date, and previous Chandra observation has shown that the strong emission lines including Si and Fe have a consistent Doppler shift \citep{marshall2002} within uncertainties smaller than the difference in velocity dispersion measured with Resolve. This result is based on a time-averaged spectrum from a certain epoch, and the possibility of a difference in jet precession and nodding motion between the Fe and Si K regions cannot be completely ruled out. However, taken together with our Fe K band time-averaged analysis, it would be natural to interpret the observed difference in the line width as mainly reflecting spatial variations of velocity dispersion within the jet, rather than originating from the variations of the Doppler shift due to the precession or nodding motion.

Moreover, modelling the time-averaged spectrum around the Fe K band, we found that the receding jet has a smaller line width than the approaching jet by $\sim 500$ km s$^{-1}$. 
As explained in Section~\ref{subsec:and_Fe_vs_Si}, the observed difference is unlikely to be attributed to the Doppler shift variation of the jet precession and nodding motion, and is instead likely due to a difference in velocity dispersion arising from another mechanism within the jet. This difference could be explained if the inner part of the receding jet is obscured over a larger region than the approaching jet. Our XRISM observation was conducted at a precession phase $\phi_{\rm jet} = 0.21$--$0.24$, when the approaching jet is relatively inclined to our line-of-sight, so the inner part of the receding jet could be widely obscured even outside the eclipse, by the accretion flow and maybe also the disk wind. The lower plasma temperature obtained from the receding jet component compared to the approaching jet is also consistent with the interpretation that the former traces the outer regions of the jet.
If this is the case, the line widths can change at different jet precession phases. In this manner, the structure of innermost parts of jets can be investigated by the profiles of the X-ray emission lines and their variation associated with the eclipses and precession (\cite{miller2014}). This would be one of the important subjects that should be tackled via future high-resolution spectroscopy of SS 433 with XRISM. 

What caused the decrease of the velocity dispersion along the jet axis? 
\citet{namiki2003} suggested that the origin is progressive jet collimation (decreasing opening angle of the jet). In this case, the observed velocity dispersions can be converted into the opening angle of the line emitting regions through the equation $\sigma'_{\rm jet} = f v_{\rm jet} \tan(\Theta)$, where $\sigma'_{\rm jet}$, $f$, $v_{\rm jet}$, and $\Theta$ are the velocity dispersion perpendicular to the jet axis (i.e., $\sigma'_{\rm jet} = \sigma_{\rm  jet} \sin \alpha$, where $\alpha$ is the inclination angle of the jet), the form factor depending on the emissivity distribution across the jet cross section, the jet velocity (0.26$c$), and the opening half-cone angle of the jet, respectively. Assuming the uniform density in the jet perpendicular to the jet axis and the constant velocity along the jet axis, $f = 0.74$ is obtained \citep{marshall2002}. Using this value, the velocity dispersions in the Fe K band during the eclipse and non-eclipse phases ($\sigma_{\rm jet, Fe} \sim 1000$ km s$^{-1}$ and $1700$ km s$^{-1}$, respectively) give $\Theta \sim 0.9^\circ$ and $1.6^\circ$, respectively.
If this interpretation is correct, the results indicate that the relativistic jets can be somehow collimated after they launched and accelerated. 
What makes the jet collimation is unclear, but it may be achieved by the funnel-shaped accretion flow and/or disk winds, or the pressure of surrounding ambient medium, as suggested by studies of jet collimation for active galactic nuclei (e.g., \cite{nakamura18}). Alternatively, the magnetic field structure may play a central role, as proposed by \citet{namiki2003}. 
Comparison of the results that we obtained with theoretical studies and numerical simulations is required to reveal its physical mechanism of the jet collimation. 

Alternative possibility is that the jet plasma is more turbulent in the hotter region than in the colder region. Recent hydrodynamical simulations of supercritical accretion flows suggests the presence of the turbulent structures around the jets (e.g., \cite{yan-fei14}). 
However, many of such simulations are focused on the vicinity of the compact object, with spatial scales 
much smaller than the location of the X-ray emitting region of the jets in SS 433, which is estimated to be $\sim 10^{12}$ cm \citep{kotani1996} from the compact object. Larger-scale simulations are required to investigate if the turbulent structure remains in outer regions of the jets and can explain the variation of the velocity dispersion that we observed. 

As a byproduct of the study of jet velocity dispersion, we were able to set a constraint on the Ni abundance, confirming it to be $6.3^{+0.5}_{-0.4}$ times the solar value when adopting the solar abundance table of \cite{lodders09} (or $ 9.4^{+0.8}_{-0.6}$ when that of \cite{wilms2000} is used). 
The Ni overabundance is consistent with previous results from other X-ray observatories \citep{kotani1998,brinkmann2005,medvedev2018}. 
Notably, the Resolve allowed us for the first time to clearly resolve the emission lines above $\sim 8$ keV, including the Ni XXVIII Ly$\alpha$, which was challenging with earlier instruments due to their limited spectral resolution and sensitivity.

In summary, the XRISM/Resolve has settled the issue on the variation of the radial velocity dispersion in the SS 433 jets. Overall, the spectra in the limited energy bands were well reproduced by the single-temperature optically-thin ionized plasma model with symmetric velocity dispersion. 
However, as shown in Fig.~\ref{fig:Si_and_Fe_fit}, we have found some residuals that cannot be explained this simple bipolar jet plasma model, especially around the redshifted Ni XXVII He$\alpha$ line 
around 6.2--7 keV. In addition to the narrow FeI K$\alpha$ line at 6.4 keV, which was seen in previous observations (e.g., \cite{kotani1996,marshall2002,medvedev2018}) and often interpreted as a  reflection on the outer disk or disk winds, two broad Gaussian lines at $\sim$ 6.4 keV and $\sim$ 6.9 keV were found to reproduce the structure. Residual structure at similar energies beyond the contribution of the narrow FeI K$\alpha$ line was found in \citet{medvedev2019}, who attributed its origin to fluorescence from the inner, more highly ionized regions of the disk wind. The broader lines observed at different energies in our analysis could be interpreted as originating from regions with different ionization states. More detailed analysis and interpretation of the narrow FeI K$\alpha$ emission line and the broad components are provided in a different paper \citep{takagi2025}. 
In addition, different plasma temperature in 2--4 keV and 5.5--9 keV clearly indicates the multi-temperature nature of the X-ray emitting region of the jet. Indeed, the jet is considered to be gradually cooled as it moves apart from the compact object \citep{brinkmann1988,kotani1996} and previous Chandra observations suggest that the observed X-ray spectra are explained by at least several different temperature components \citep{marshall2013}. These demonstrate the limitations of our spectral modeling, and needs for broad-band spectral analysis with more complex structures of plasma properties in the jet.
Such complexity in the jet emission model and contributions from the accretion disk and disk winds may also affect the measurement of elemental abundances, as inaccurate assumptions about the plasma structure can lead to systematic biases in the inferred line strengths and abundance ratios.
More detailed modeling using the XRISM/Resolve data in the entire energy range would be therefore valuable for revealing the plasma properties and determining the abundances precisely, which we leave as a future work.


\begin{ack}
MS thanks Dr. Shin Mineshige, Dr. Taichi Igarashi, and Dr. Mami Machida for discussions from the viewpoint of theory and simulations. This research has made use of software provided by the High Energy Astrophysics Science Archive Research Center (HEASARC), which is a service of the Astrophysics Science Division at NASA/GSFC. Part of this work was financially supported by Grants-in-Aid for Scientific Research 19K14762, 23K03459, 24H01812 (MS), and 20H01946 (YU) from the Ministry of Education, Culture, Sports, Science and Technology (MEXT) of Japan. MS and TT acknowledge support by Ehime University Grant-in-Aid Research Empowerment Program. 
\end{ack}



\bibliographystyle{apj}
\bibliography{ss433}{}

\appendix 





\end{document}